\documentclass[12pt,preprint]{aastex}

\usepackage{epsfig}
\usepackage{amsmath}
\usepackage{natbib}

\pdfoutput=1

\newcommand\Spitzer{\textit{Spitzer }}

\shorttitle{Mid-Infrared Spectroscopy of Two Lensed Star-forming Galaxies}

\shortauthors{Fadely et al.}

\begin{document}

\title{Mid-Infrared Spectroscopy of Two Lensed Star-forming Galaxies}
\author{Ross Fadely\altaffilmark{1}, Sahar S. Allam\altaffilmark{2}, 
 Andrew J. Baker\altaffilmark{1}, \\
 Huan Lin\altaffilmark{2}, Dieter Lutz\altaffilmark{3}, Alice E. 
Shapley\altaffilmark{4},  Min-Su Shin\altaffilmark{5}, \\
J. Allyn Smith\altaffilmark{6}, Michael A. Strauss\altaffilmark{7}, \& 
Douglas L. Tucker\altaffilmark{2}}

\altaffiltext{1}{Department of Physics and Astronomy, Rutgers, the State 
University of New Jersey, 136 Frelinghuysen Road, Piscataway, NJ 08854-8019;
\{fadely,ajbaker\}@physics.rutgers.edu}

\altaffiltext{2}{Fermi National Accelerator Laboratory, P.O. Box 500, Batavia, 
IL 60510; \{sallam,hlin,dtucker\}@fnal.gov}

\altaffiltext{3}{Max-Planck-Institut f\"ur extraterrestrische Physik, 
Postfach 1312, D-85741 Garching, Germany; lutz@mpe.mpg.de}

\altaffiltext{4}{Department of Astronomy, University of California, Los 
Angeles, 430 Portola Plaza, Los Angeles, CA 90024; aes@astro.ucla.edu}

\altaffiltext{5}{Department of Astronomy, University of Michigan, 500 Church 
Street, Ann Arbor, MI 48109; msshin@umich.edu}

\altaffiltext{6}{Department of Physics and Astronomy, Austin Peay State 
University, P.O. Box 4608, Clarksville, Tennessee 37044; smithj@apsu.edu}

\altaffiltext{7}{Department of Astrophysical Sciences, 4 Ivy Lane, Peyton Hall,
Princeton University, Princeton, NJ 08544; strauss@astro.princeton.edu}

\begin{abstract}
We present low-resolution, rest-frame $\sim 5-12$\,$\mu$m $Spitzer$/IRS 
spectra of two lensed $z \sim 2$ UV-bright star-forming galaxies, 
SDSS\,J120602.09+514229.5 and SDSS\,J090122.37+181432.3.  Using the 
magnification boost from lensing, we are able to study the physical properties 
of these objects in greater detail than is possible for unlensed systems.  In 
both targets, we detect strong PAH emission at 6.2, 7.7, and 11.3\,$\mu$m, 
indicating the presence of vigorous star formation.  For J1206, we find a 
steeply rising continuum and significant [\ion{S}{4}] emission, suggesting that 
a moderately hard radiation field is powering continuum emission from small 
dust grains.  The strength of the [\ion{S}{4}] emission also implies a  
sub-solar metallicity of $\sim 0.5\,Z_\odot$, confirming published rest-frame 
optical measurements.  In J0901, the PAH lines have large rest-frame 
equivalent widths ($>1$\,$\mu$m) and the continuum rises slowly with 
wavelength, suggesting that any AGN contribution to $L_{\rm{IR}}$ is 
insignificant, in contrast to the implications of optical emission-line diagnostics.  
Using [\ion{O}{3}] line flux as a proxy for AGN strength, we estimate that the AGN 
in J0901 provides only a small fraction of its mid-infrared continuum flux.  
By combining the detection of [\ion{Ar}{2}] with an upper limit on [\ion{Ar}{3}] 
emission, we infer a metallicity of $\gtrsim 1.3\,Z_\odot$.
This work highlights the importance of combining rest-frame optical and mid-IR 
spectroscopy in order to understand the detailed properties of star-forming 
galaxies at high redshift. 

\end{abstract}

\section{Introduction}

Rest-frame UV selection offers a prime view into populations of star-forming
galaxies at $z > 1.5$.  First used to identify $z \sim 3$ galaxies with sharp
breaks in their spectral energy distributions due to absorption below the 
Lyman limit \citep{1996AJ....112..352S}, the Lyman break technique has now 
been modified and extended to both lower and higher redshift 
\citep[e.g.,][]{1999ApJ...519....1S,2003ApJ...593..630L,2004ApJ...607..226A}.
Since the development of the technique in the 1990s, thousands of star-forming galaxies 
have been spectroscopically confirmed at $z \sim 1.5 - 3.5$
\citep[e.g.,][]{2003ApJ...592..728S,2006ApJ...653.1004R}, resulting in a 
revolution in our understanding of galaxy formation and evolution.  However,
these high-redshift galaxies suffer from a fundamental problem: they are 
typically small and faint, with $R_{\rm AB} \geq 24$\,mag, making it 
impossible to carry out detailed studies of individual objects unless they 
happen to be strongly lensed.  In particular, this limitation applies to observations 
of UV-bright star-forming systems at the long wavelengths that can trace star 
formation even in highly 
obscured regions.  While stacking analyses of large samples 
confirm that UV-selected galaxies at $z \sim 2$ have substantial 
fractions of their bolometric luminosities emerging in the far-infrared, 
with $\left<L_{\rm IR}/L_{\rm UV}\right> \simeq 4-5$ 
\citep{2004ApJ...603L..13R,2006ApJ...644..792R,2010ApJ...712.1070R}, understanding the 
parameters of obscured star formation in individual unlensed objects remains 
out of reach for current facilities.

The first bright lensed Lyman break galaxies, ``cB58'' (i.e., MS\,1512-cB58)
and the ``Cosmic Eye'' (i.e., LBG\,J213512.73$-$010143), were discovered 
serendipitously in the course of a cluster redshift survey 
\citep{1996AJ....111.1783Y} and a {\it Hubble Space Telescope} ({\it HST}) 
snapshot imaging survey of X-ray-bright clusters \citep{2007ApJ...654L..33S}, 
respectively.  Since clusters are rich in strong caustics capable of 
producing fold arcs (like cB58) and in individual galaxies capable of producing 
nearly complete Einstein rings (like the Cosmic Eye), the circumstances of 
these discoveries were not surprising.  More recently, however, several teams
have begun to exploit the enormous footprint of the Sloan Digital Sky 
Survey (SDSS) to identify UV-bright high-redshift sources that are lensed by 
individual galaxies in field or group environments.  This enterprise
kicked off with the serendipitous discovery of the ``8 O'Clock Arc'' 
\citep{2007ApJ...662L..51A} and has now spawned a variety of systematic 
searches within the SDSS object catalog that rely on different selection 
criteria \citep[e.g.,][]{2007ApJ...671L...9B,2008AJ....136...44S,2008AJ....135..664H}.  By 
focusing on luminous red galaxies with 
multiple blue neighbors, and on close pairs with characteristic 
lens+arc morphologies, various authors of this paper have now 
contributed to the discovery of 11 new spectroscopically confirmed lenses at 
redshifts $0.4 \leq z \leq 2.4$ 
\citep{2009ApJ...696L..61K,2009ApJ...699.1242L,2009ApJ...707..686D}.

In this paper, we present {\it Spitzer}/IRS spectroscopy of two objects from 
this new SDSS sample.  The first, SDSS\,J120602.09+514229.5 (a.k.a. the ``Clone'', 
hereafter J1206) is a $z = 2.00$ arc discovered by \citet{2009ApJ...699.1242L},
who determine a lensing magnification ${\cal M} = 27 \pm 1$.  The second, 
SDSS\,J090122.37+181432.3 (hereafter J0901) is a $z = 2.26$ arc discovered by 
\citet{2009ApJ...707..686D}; preliminary lens modelling implies a 
magnification $\approx 8$ (A. West et al. 2010 in preparation).  
Both objects are 20--30 times brighter than galaxies at the knee 
of the $1.9 \leq z \leq 2.7$ rest-UV luminosity function 
\citep{2008ApJS..175...48R}; their estimated intrinsic far-IR luminosities 
make J1206 a luminous infrared galaxy (LIRG; see Section \ref{sec:1206})
and J0901 an ultraluminous infrared galaxy (ULIRG; see Section \ref{sec:0901}).
Here we focus on what can be learned about the conditions in 
the dusty regions of these galaxies from their integrated mid-infrared 
emission, based on comparisons with local galaxies, assuming that there is 
minimal differential lensing across the wavelength ranges of our IRS spectra.  
We defer to future papers analysis of {\it 
Spitzer}/IRAC and (for J1206) MIPS imaging of these targets in light of 
more refined {\it HST}-based lens models, together with detailed comparisons 
to the source-plane properties of similar lensed star-forming galaxies \citep{2008ApJ...689...59S,2009ApJ...698.1273S}
and high-redshift systems selected through their dust emission rather than rest-UV colors \citep[e.g.,][]{2005ApJ...625L..83L,
2006ApJ...638..613W,2007ApJ...660.1060V,2007ApJ...658..778Y,2008ApJ...675.1171P,2009ApJ...699..667M}.

\section{Observations and Data Reduction}
\label{sec:obs}

We used the Infrared Spectrograph \citep[IRS:][]{2004ApJS..154...18H} on board the 
\textit{Spitzer Space Telescope} to obtain $14-38$\,$\mu$m spectra of both 
J1206 and J0901 in the instrument's ``long low'' mode ($R \sim 57 - 126$), for 
which the first (LL1) and second (LL2) orders cover wavelength ranges of 
$19 - 38$\,$\mu$m and $14 - 21.3$\,$\mu$m, respectively.  In order to ensure 
optimal signal-to-noise ratios, we followed the recommendations of 
\citet{2007ApJ...659..941T} and mapped the targets at six positions across 
the slit.  Observations of J1206 were taken during \Spitzer Cycle 4 (PID 
40430; PI S. Allam) on 2007 December 19--20, and consisted of $1 \times 6$ 
pointings in LL1 and LL2 for total integration times of 2.1\,ks and 2.2\,ks, 
respectively.  J0901 was observed on 2009 May 15 during \Spitzer Cycle 5 
(PID 50086; PI S. Allam) using $2 \times 6$ pointings in LL1 totaling 2.2\,ks 
and 2.0\,ks, and $1 \times 6$ pointings in LL1 totaling 2.2\,ks.  Data 
were obtained under nominal operating conditions, with the exception of the 
second LL1 Astronomical Observation Request (AOR) for J0901.  During this AOR, 
\Spitzer began to warm up due to the depletion of its cryogen.  The increased 
thermal background was marginal, raising data collection event (DCE) values by 
only 6\%.  The IRS support team deemed the data nominal, and we reduced them 
following the same procedures used for our other AORs.

Data reduction relied on standard analysis packages and followed the procedure 
described by \citet{2007ApJ...659..941T}.  Using IRAF, we removed latent 
charge in the IRS images row-by-row by fitting the linear background increase 
over time.  Subsequently, we masked ``rogue'' pixels using the IDL routine 
IRSCLEAN.  After cleaning, we constructed sky images for each target position 
using the five other pointings from the same mapping AOR.  The resulting sky 
images were subtracted from the corresponding science frames, and the 
differences were then co-added to produce a final 2D spectrum at each 
position.  We extracted 1D spectra using the SPICE package provided by the 
Spitzer Science Center, using optimal extraction.  We used an extraction 
aperture half the default size in order to avoid contamination from other 
sources (see below), but we corrected to full-aperture fluxes using 
observations of standard stars.

\begin{figure}[h]
\centering
\includegraphics[clip=true, trim=9.15cm 1.cm 11.8cm 6.50cm,width=8cm]{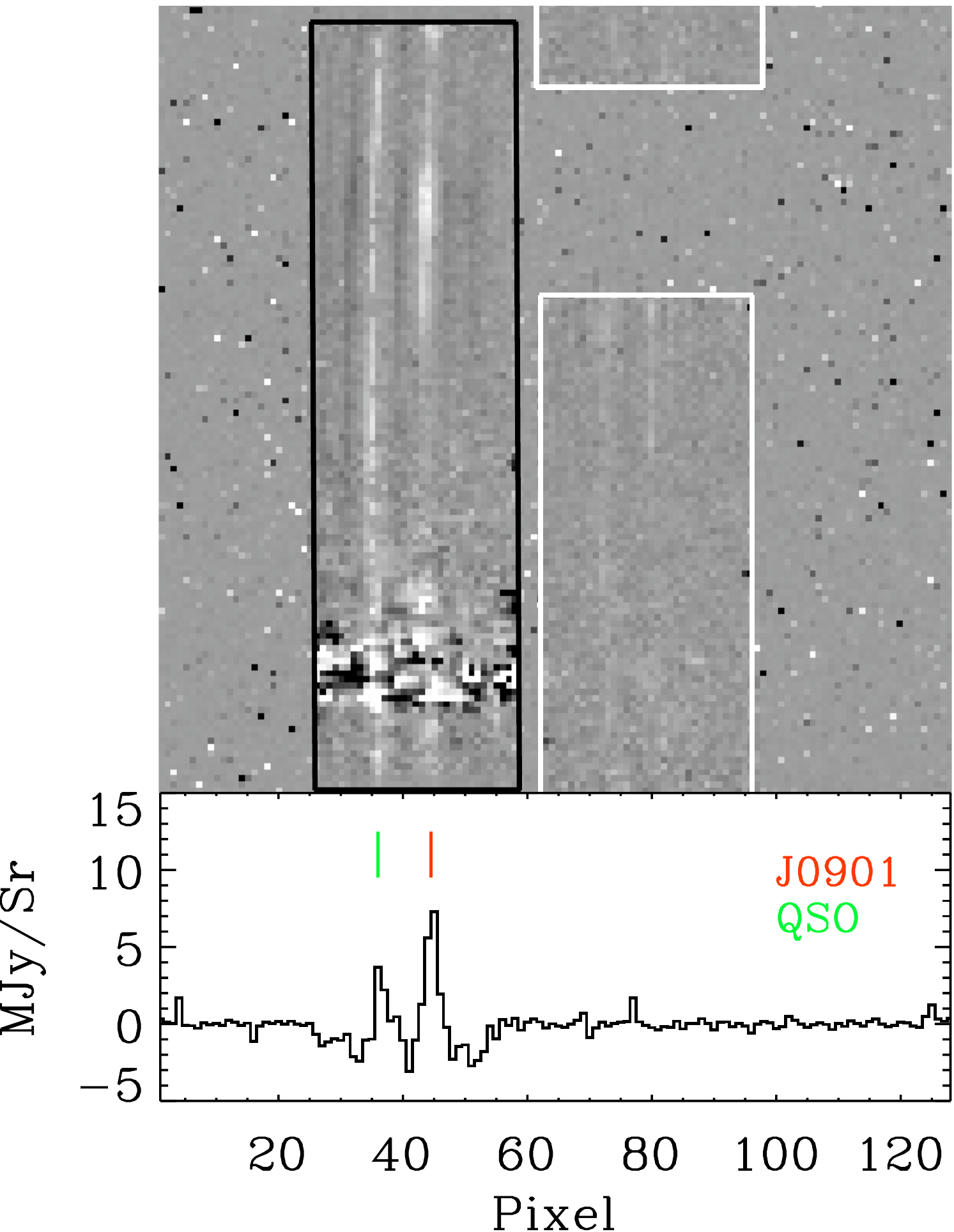}
\includegraphics[clip=true, trim=9.15cm 1.cm 11.8cm 6.50cm,width=8cm]{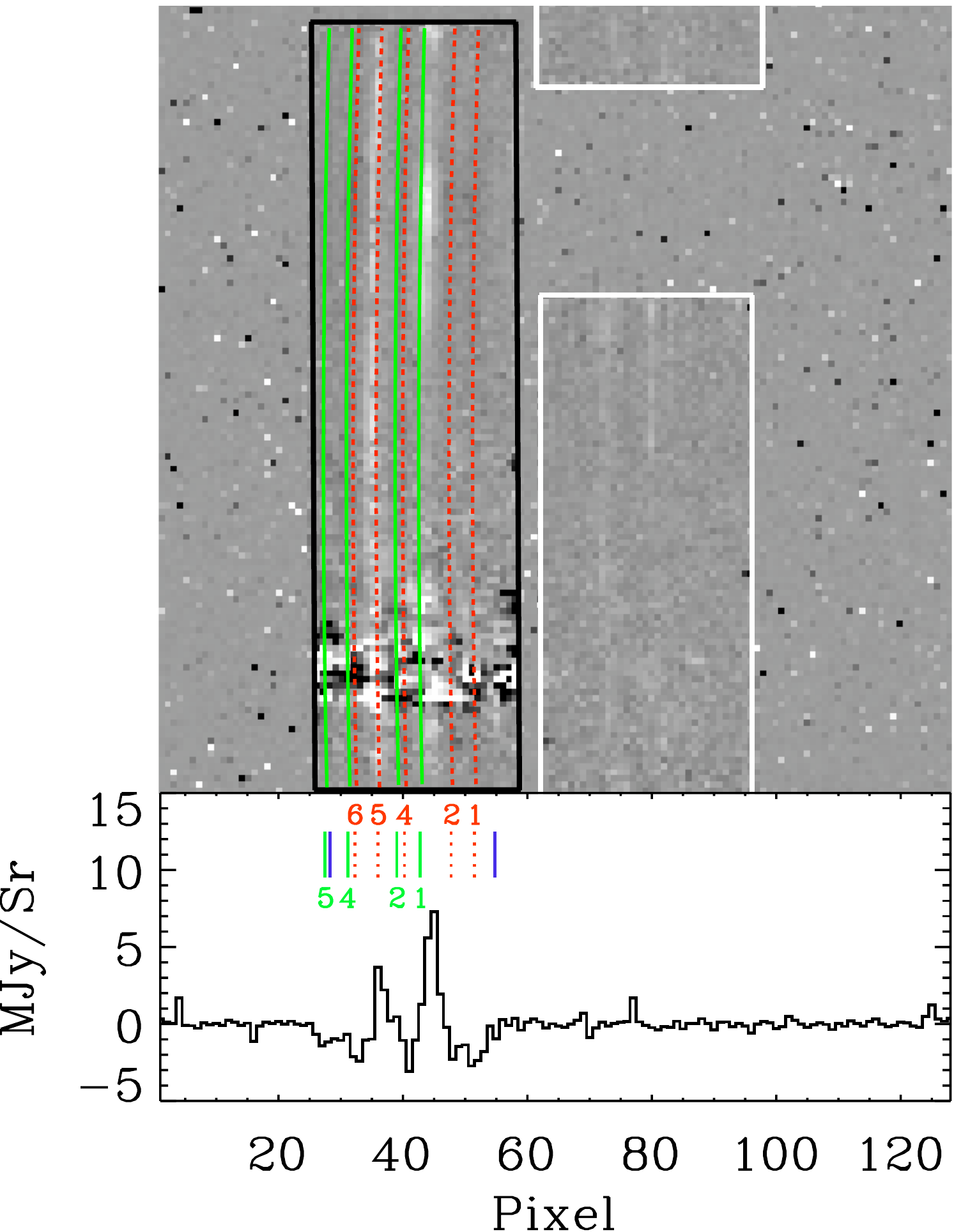}
\caption{(a) Left: reduced 2D spectrum for J0901 at our third map position, after 
subtraction of a model for the sky.  The black and white rectangles indicate 
the areas of the array exposed to the spectrometer for LL1 and LL2, 
respectively, with wavelength increasing downward.  Two LL1 spectra are 
visible: the detection on the right is our target J0901, while on the left is 
the interloping source SDSS\,J090125.59+181427.8.  From the pattern of the 
neighboring pixels, it is clear that the background is oversubtracted due to 
the relatively bright spectra of J0901 and the additional source.  Plotted 
below the spectrum are the data for the row corresponding to the peak of the 
J0901 spectrum, where the oversubtraction is the worst.  (b) Right: same 
data overplotted with the positions of the two spectra for the other map 
positions, indicated by the numbers in the lower panel.  The dotted red lines 
mark the positions of the J0901 spectra, while the solid green lines mark 
those of the additional source.  For this (i.e., the third) map position, the 
spectrum for J0901 lies just next to the rightmost solid green line (the 
position of the interloping source for our first map position) and between two 
dashed red lines (the positions of J0901 for our second and fourth map 
positions). To correct for this effect, we measure the sky at appropriate 
positions, indicated by the blue line segments.  
The combination of these measurements gives an accurate model for the 
oversubtraction, which is then added to the data.}
\label{fig:0901data}
\end{figure}

For J0901, extraction of the 1D spectra was complicated by an interloping 
source, SDSS\,J090125.59+181427.8 (see Figure \ref{fig:0901data}a), lying 
$\sim 46''$ away from the lens.  From its IRS spectrum, this object is likely to be a quasar at $z \sim 1.3$, in 
agreement with the assessment of its optical colors by \citet{2009ApJS..180...67R}.  Its spectrum in one 
pointing often lay on or near the position of a J0901 spectrum in {\it 
another} pointing; combined with the fact that J0901 itself is bright and extended, 
this situation meant that a given ``sky'' frame included the true background, 
light from the interloping source, and residual flux from J0901 at other 
map positions, leading to oversubtraction in our final 2D spectra (e.g., 
Figure \ref{fig:0901data}a).  To correct for this effect, we 
extracted 1D spectra of the negative sky echoes at {\it different} locations 
in the 2D spectrum whose combination should have experienced the {\it same} 
oversubtraction as the position in question (e.g., Figure \ref{fig:0901data}b).
By combining such measurements, we constructed an empirical model for 
the 
oversubtraction of each of our target spectra that accounted for both 
continuum and PAH features.  We found 
that the correction was fairly mild, leading to a $\sim 10-25$\% flux increase 
over the uncorrected spectrum.  

As an alternative to the above procedure, we tried to create sky images using 
only dithers whose positions were far enough away that no oversubtraction 
occurred.  Doing so meant constructing sky images from many fewer frames, 
often from only one or two other positions.  As a result, we found the reduced 
(cleaned and sigma-clipped) 2D spectra were affected by rogue pixels whose
effects would otherwise have been eliminated.  Such pixels not only increased 
the uncertainty in our final 1D spectrum, but also introduced small-scale 
features that do not correspond with any known emission features.  Given the 
undesirable effects of this approach to sky subtraction, we opted for the 
method described above, which uses all of the original frames.  We note that 
the spectra produced via our two sky subtraction procedures yield similar flux 
densities within the measurement uncertainties.  

For both J1206 and J0901, the error spectrum at each position was calculated 
using SPICE and standard deviation frames constructed in IRAF.  The final 
error spectrum is that of all the positions added in quadrature.  For J0901, 
we include an additional systematic uncertainty associated with our sky 
correction, raising the final error spectrum by a factor $\sim \sqrt{2}$.

\section{Results}
\label{sec:results}

\subsection{SDSS\,J120602.09+514229.5}
\label{sec:1206}

\begin{figure}[h]
\centering
\includegraphics[clip=true, trim=3.1cm 13.05cm 2.5cm 2.75cm,width=8cm]{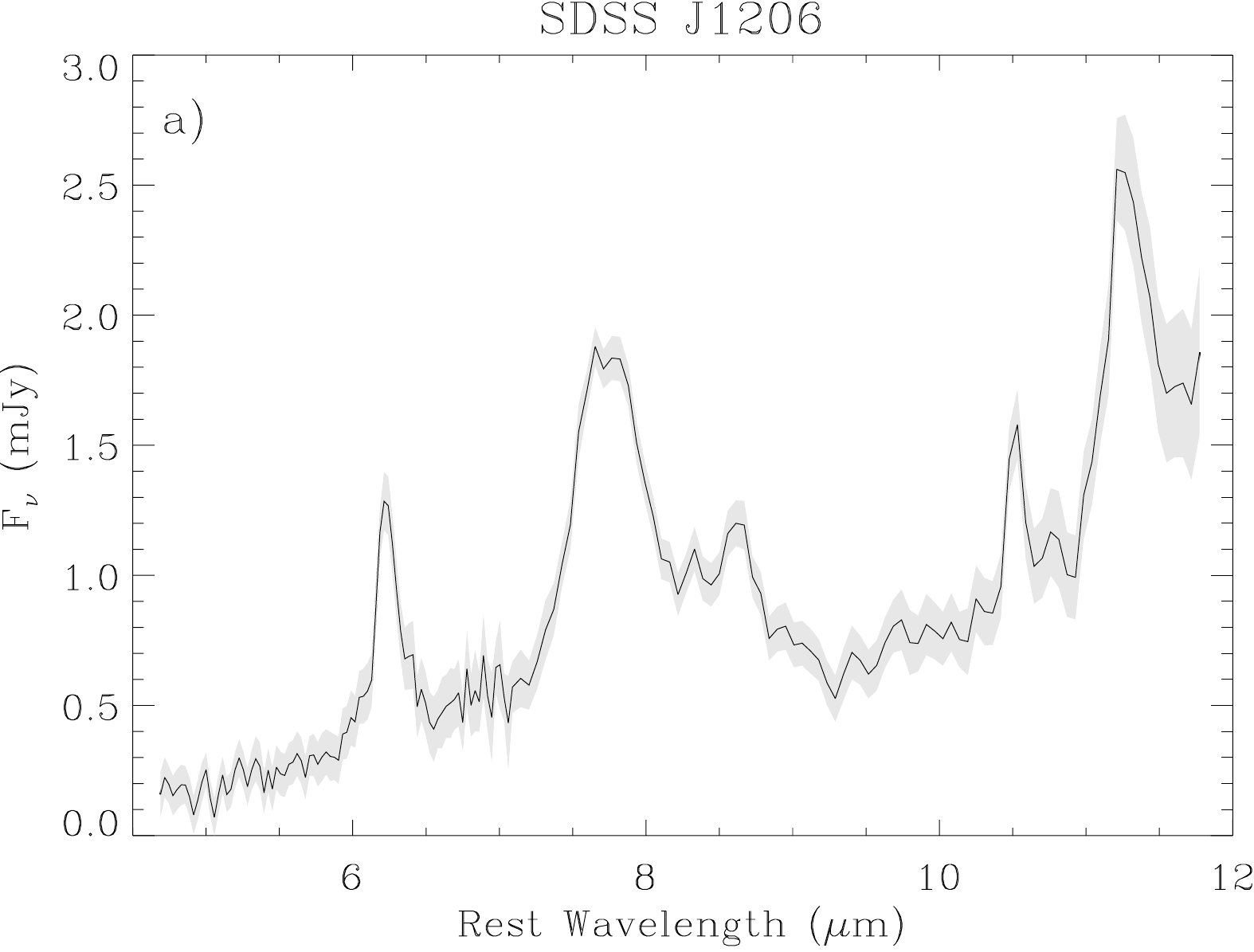}
\includegraphics[clip=true, trim=3.1cm 13.05cm 2.5cm 2.75cm,width=8cm]{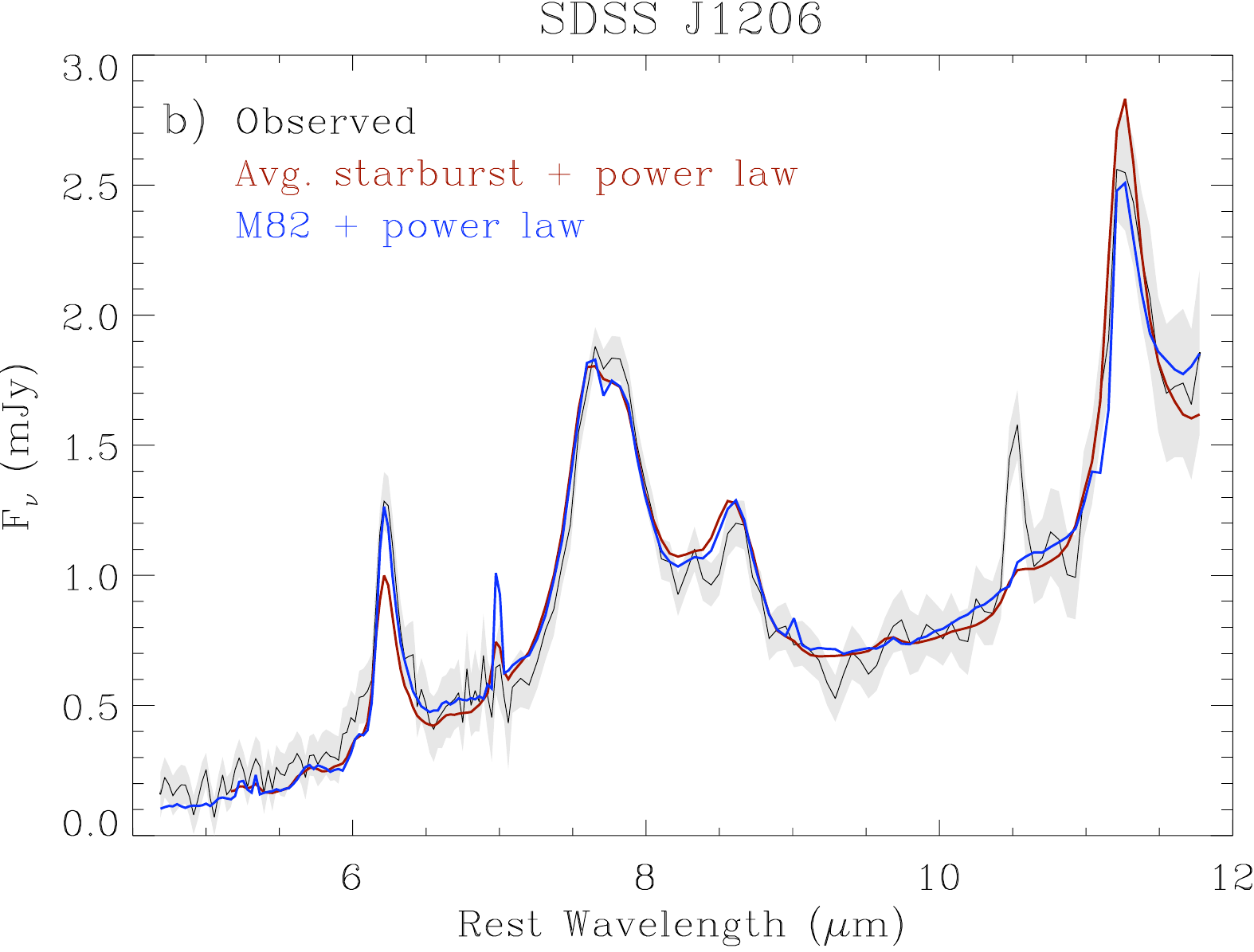}
\includegraphics[clip=true, trim=3.1cm 13.05cm 2.5cm 2.75cm,width=8cm]{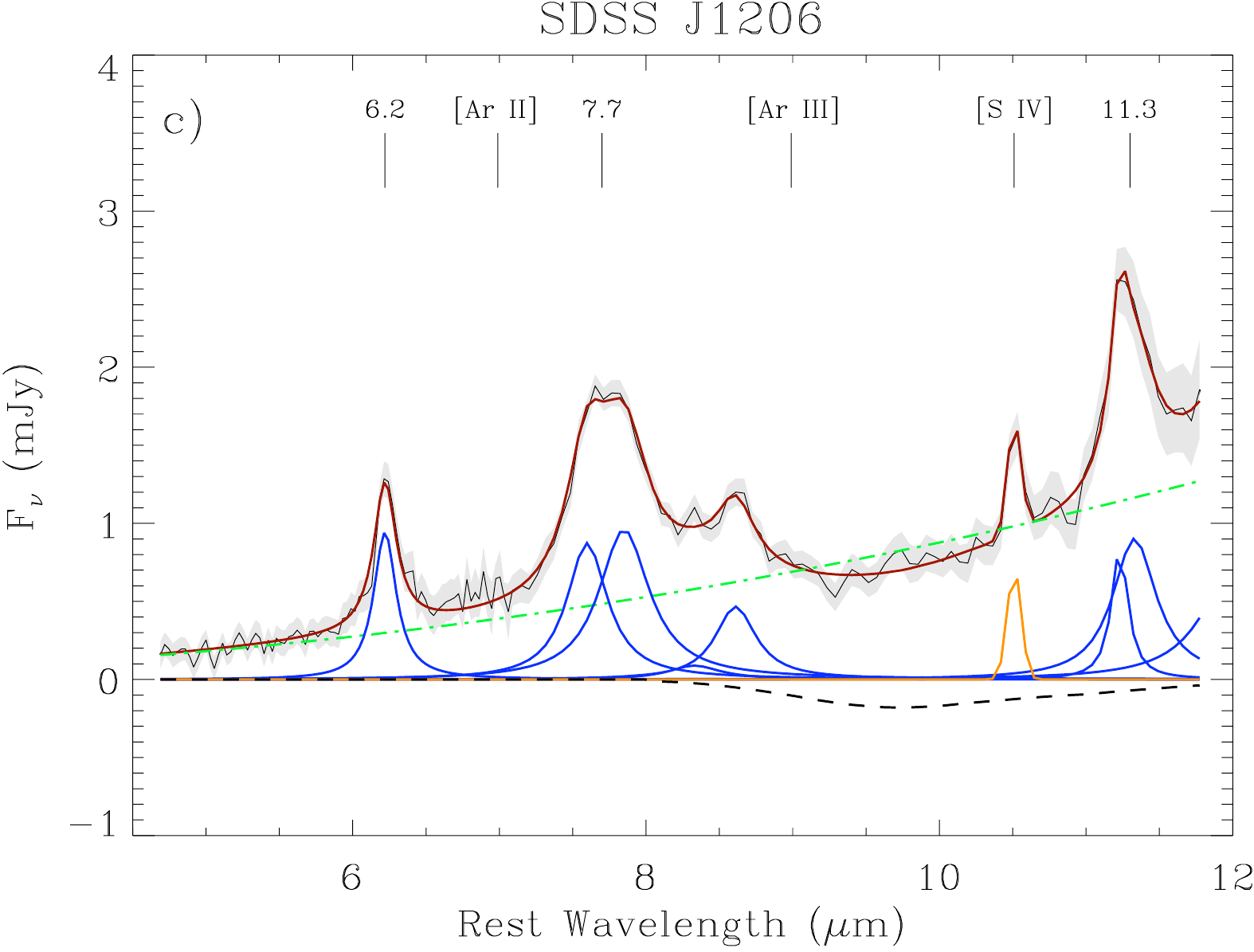}
\caption{(a) IRS spectrum of J1206.  The observed spectrum is plotted in 
black after redshifting to rest wavelength for $z = 2.00$.  The light 
grey band corresponds to the $1\sigma$ uncertainty associated with the 
spectrum.  (b)  Plotted in blue and red, respectively, are the best fit template + power law spectra 
(see text) for M82 
and the average starburst spectrum of Brandl et al. (2006) for the parameters 
listed in Table \ref{tab:templates}. (c) A comprehensive fit (red) to the 
spectrum using Drude profiles for PAH features (blue), Gaussian profiles for 
ionic lines (orange), and a power law continuum (green dot dashed).  
The fit also includes 9.7 $\mu$m silicate extinction (black dashed).}
\label{fig:1206}
\end{figure}

Figure \ref{fig:1206}a shows the extracted IRS spectra for J1206.  On 
top of a rising continuum, prominent PAH features are present at 6.2, 
7.7, and 11.3\,$\mu$m.  In addition, a strong [\ion{S}{4}] feature is 
present at 10.5\,$\mu$m.   In order to compare the spectrum to those of local 
starbursting analogs, we fit template \textit{Infrared Space Observatory} 
spectra of 
30 Doradus, Circinus, M82, NGC 253, and NGC 1068 
from \citet{2000A&A...358..481S}, and the 
average \textit{Spitzer} starburst template from \citet{2006ApJ...653.1129B}.  
For the fit we allow a varying contribution from the templates as well as an 
additional power law continuum (normalized to 6.2\,$\mu$m): $F_{\nu,{\rm fit}} 
= C_1 \times [{\rm Template}]+ C_2 \times 
(\lambda/6.2\,\mu{\rm m})^\alpha$, where $C_1$ is dimensionless and $C_2$ has 
units of mJy.  The three parameters for the fit 
were sampled using a standard Metropolis-Hastings MCMC (Markov chain 
Monte Carlo) algorithm.  Figure \ref{fig:1206}b shows two of the template 
spectra that provide good fits.  Table \ref{tab:templates} shows the inferred 
median, 68\% confidence limits, and best-fit parameters 
for the templates that provided reasonable fits.
We find that the spectrum is well fit by the M82 template ($\chi^2_{\rm red} 
= 0.96$), and marginally fit by the NGC\,253 and average starburst templates 
($\chi^2_{\rm red}=1.20,1.14$).  In contrast, the other templates from 
\citeauthor{2000A&A...358..481S} (for 30 Doradus, Circinus, and NGC\,1068) 
fit the spectrum poorly ($\chi^2_{\rm red}>10$) due to the lack of strong PAH 
features.  For each of the acceptable 
templates, the best fit includes an additional, steeply-rising continuum 
with a power law index of $\alpha \sim 3.3$.  

Exploiting the similarity to the spectrum of M82, we can estimate the far-IR 
luminosity of J1206 as  
\begin{eqnarray}
L_{\rm FIR}({\rm J1206}) & = & L_{\rm FIR}({\rm M82}) \cdot \Big({\frac 
{D_{L}({\rm J1206})}{D({\rm M82})}}\Big)^2 \cdot {\frac 1{\cal M}} \cdot 
{\frac {S_{\nu^\prime}({\rm J1206})\,\Delta \nu^\prime}
{S_{\nu}({\rm M82})\,\Delta \nu}} \\
 & = & 2.76 \times 10^{16}\,L_\odot \cdot {\frac 
{S_{\nu^\prime}({\rm Clone})\,\Delta \nu^\prime}
{S_{\nu}({\rm M82})\,\Delta \nu}} \label{e-lfir2}
\end{eqnarray}
in terms of the distance to M82 \citep[3.63\,Mpc: ][]{2001ApJ...553...47F} 
and the corresponding $40-400\,{\rm \mu m}$ far-IR luminosity 
\citep[$4.1 \times 10^{10}\,L_\odot$: ][]{2003AJ....126.1607S}, the 
magnification of J1206 \citep[${\cal M} = 27 \pm 1$: ][]{2009ApJ...699.1242L},
and the luminosity distance to J1206 for a $(h,\Omega_m,\Omega_\Lambda) = 
(0.7,0.3,0.7)$ cosmology.  The remaining term in Equation \ref{e-lfir2} is the 
ratio of the observed spectra in flux density units after corresponding 
rest-frequency channels ($\Delta \nu$ and $\Delta \nu^\prime=\Delta\nu/(1+z)$) are matched up.
This is precisely the value of $C_1$ 
delivered by our template fitting, leading to an estimate of $L_{\rm FIR} = 
2.4 \times 10^{11}\,L_\odot$ that places J1206 in the LIRG regime.  In estimating 
$L_{\rm FIR}$ we ignore the additional power-law component used in template 
fitting, since its contribution should be negligible at far-IR wavelengths.

To extract and interpret the features of the spectrum, we have also computed a 
fit using the Drude profiles defined in \citet{2007ApJ...657..810D} for 
PAH features, Gaussian line profiles for ionized species expected to be strong, and a 
simple power law continuum ($\propto \lambda^{\alpha}$).  In addition, the fit
includes extinction effects from 9.7 $\mu$m silicate absorption.  The results of this 
fit are plotted in Figure \ref{fig:1206}c and tabulated in Table 
\ref{tab:results}.  To assess the strength of the PAH emission, we compare 
the rest-frame equivalent widths (EWs) of the 6.2, 7.7 ($\equiv 7.41+7.61+7.85$), and 
11.3 ($\equiv 11.23+11.33$) $\mu$m PAH features to those of local starbursting 
systems.  Comparisons of EWs are known to be sensitive to the details of 
how authors define the underlying continuum.  In particular, continuum levels
are often defined by the values of the data seen on either side of emission features 
\citep[see e.g.,][]{2006ApJ...653.1129B,2008ApJ...675.1171P}.  Such definitions result in systematically higher 
continua and lower PAH EWs than found by comprehensive 
fits to the spectra \citep[e.g.,][]{2009ApJ...698.1273S}.  To facilitate 
interpretation of the EW values reported in Table \ref{tab:results}, we analyze
the average starburst spectrum of \citet{2006ApJ...653.1129B} using the same
conventions as for J1206.  We find EWs for the average starburst spectrum that 
are factors $\sim 1-8$ higher than derived by \citet{2006ApJ...653.1129B} for 
their own data.  Nevertheless, this approach provides a consistent way of 
measuring and comparing EW values to our data.  Relative to the average 
starburst spectrum, J1206 has EWs that are lower by factors of 1.2, 1.5, and 
1.2 for the 6.2, 7.7, and 11.3\,$\mu$m PAH features, respectively.  This 
slight deficiency is not surprising, given the additional power-law continuum
 preferred by our template fitting above.

The two most striking features of the $\sim 4.5-12$\,$\mu$m spectrum of J1206 
are its steep underlying continuum and prominent [\ion{S}{4}] emission.  In 
high-resolution studies of local star-forming galaxies, the latter line is 
fairly common and appears weakly in starbursts \citep{2009ApJS..184..230B} and 
ULIRGs  
\citep[][]{2007ApJ...667..149F}, but is much stronger in blue compact 
dwarfs \citep[BCDs;][]{2009ApJ...704.1159H}.  At lower resolution this line is 
unresolved, and only the strongest emitters are detected \citep[see, 
e.g.,][]{2006ApJ...653.1129B, 2006ApJ...639..157W}.  Comparing the relative 
strengths of [\ion{S}{4}] and PAH emission of J1206 to those of local 
counterparts, we identify the two low-resolution mid-IR spectra of the 
starburst NGC\,1222 \citep{2006ApJ...653.1129B} and the BCD UGC\,4274 
\citep{2006ApJ...639..157W} as close analogs.

Using the optical spectroscopy of \citet{1995ApJ...450..547L} for NGC\,1222 
and \citet{1997ApJS..112..391H} for UGC\,4274, we calculate the oxygen 
abundance in each galaxy using the $N2$ and $O3N2$ indicators calibrated by  
\citet{2004MNRAS.348L..59P}.  We find $12+\log(\rm{O/H})_{N2}=8.48,8.43$ and 
$12+\log(\rm{O/H})_{O3N2}=8.38,8.37$ for NGC\,1222 and UGC\,4274, 
respectively.  For J1206, \citet{2009ApJ...701...52H} use near-IR 
spectroscopy to find $12+\log(\rm{O/H})_{N2}=8.50\pm0.18$ and 
$12+\log(\rm{O/H})_{O3N2}=8.34\pm0.14$, in good agreement with the putative 
local counterparts.  Given the consistency between the spectra and the oxygen 
abundances, modulo a factor $\sim 2.5$ uncertainty in the calibration of the 
optical diagnostics \citep{2004MNRAS.348L..59P}, we conclude that NGC\,1222 
and UGC\,4274 have physical conditions similar to those of J1206.  

In addition to oxygen abundances, we can consider the ratios of ionized sulfur 
and neon lines [\ion{S}{4}]\,10.5\,$\mu$m/[\ion{S}{3}]\,18.7\,$\mu$m and 
[\ion{Ne}{3}]\,15.6\,$\mu$m/[\ion{Ne}{2}]\,12.8\,$\mu$m for 
NGC\,1222 and UGC\,4274.  These quantities are well-known proxies for 
the hardness of the radiation field \citep[see Figure 9 
of][]{2009ApJ...704.1159H}, which is also a function of metallicity 
\citep{2006ApJ...639..157W}.  The galaxies have similar ratios of
[\ion{Ne}{3}]/[\ion{Ne}{2}] $\sim1.30$ and [\ion{S}{4}]/[\ion{S}{3}] $\sim
0.35$, and lie between lower excitation
starbursts and higher excitation BCDs on a [\ion{S}{4}]/[\ion{S}{3}]
versus [\ion{Ne}{3}]/[\ion{Ne}{2}] excitation diagram \citep{2009ApJ...704.1159H}, implying 
a moderately hard radiation field.  Using stellar models, \citet{2000ApJ...539..641T} estimate the hardness of the 
radiation in starbursts by relating 
[\ion{Ne}{3}]/[\ion{Ne}{2}] to the ratio of the infrared and Lyman continuum luminosities 
($L_{\rm IR}/L_{\rm Lyc}$).  Since NGC\,1222 and UGC\,4274 have higher
 [\ion{Ne}{3}]/[\ion{Ne}{2}] ratios than the \citeauthor{2000ApJ...539..641T} sample, we extrapolate
 their results and find $3 \lesssim L_{\rm IR}/L_{\rm Lyc} \lesssim 20$ indicating a somewhat 
 lower range than for their more typical starbursts ($4 \lesssim L_{\rm IR}/L_{\rm Lyc} \lesssim 30$).  
 If present in J1206, 
such hard radiation would cause significant heating of small dust grains and 
naturally explain the steep continuum in our IRS spectrum.  Unfortunately, 
[\ion{Ne}{3}], [\ion{Ne}{2}], and [\ion{S}{3}] all lie outside of our spectral 
coverage, preventing definitive confirmation of this hypothesis.

In J1206, the apparent consistency between metallicity estimates based on 
rest-frame optical and (via analogy with NGC\,1222 and UGC\,4274) infrared 
observations suggests that we are not seeing discrepancies of the sort seen in 
the most violent local mergers.
For ULIRGs, abundances derived from optical line diagnostics (even after 
careful extinction corrections) are lower than those derived from mid-infrared 
spectra, likely because the former are depressed by inflows of metal-poor gas 
from the outskirts of progenitor disks \citep{2008ApJ...674..172R} while the 
latter reflect rapid local enrichment in the most deeply embedded star-forming 
regions \citep{2009ApJS..182..628V}.  
Metallicities for J1206 appear internally consistent across both obscured and 
unobscured regions, suggesting a less traumatic recent history that is 
consistent with the system's relatively modest far-IR luminosity.

Our conclusions about the implied metallicity and excitation state of J1206
require a certain degree of caution.  While in good agreement with optical 
measurements, our selection of local infrared analogs for J1206 is based 
mostly on our detection of a single line, [\ion{S}{4}], whose correlations 
with PAH strength, metallicity, and spectral hardness may be uncertain by 
30\% or more 
\citep{2006ApJ...639..157W, 2009ApJ...704.1159H, 2010ApJ...712..164H}.  
Further, metallicities derived from the optical emission lines in NGC\,1222 
and UGC\,4274 are themselves uncertain by a factor $\sim 2.5$ 
\citep{2004MNRAS.348L..59P}.  Nevertheless, the consistency (within the 
measured uncertainties) between the IR spectra of J1206, NGC\,1222, and 
UGC\,4274, as well as the optical and IR properties of NGC\,1222 and 
UGC\,4274, is encouraging, indicating that the three objects do manifest 
similar physical conditions.  In the future, more robust estimates of the 
metallicity and excitation properties of J1206 will be possible with 
measurements of additional emission lines over a wider spectral range.

We can also consider the absence of other emission lines in our IRS 
spectrum,  which shows no significant ionic or molecular features 
other than [\ion{S}{4}].  In Table \ref{tab:results}, we report $3\sigma$ upper limits on [\ion{Ar}{2}] at 6.99\,$\mu$m, 
[\ion{Ar}{3}] at 8.99\,$\mu$m, and [\ion{Ne}{6}] at 7.65\,$\mu$m.
 At first glance, it seems surprising we do not detect [\ion{Ar}{2}], given 
that it is commonly detected in starburst systems \citep{2006ApJ...653.1129B}, 
including M82 and NGC\,253.  As discussed above, however, the presence of a 
rising continuum and [\ion{S}{4}] emission imply higher 
excitation in J1206 than in average starbursting systems.  
Given that  the ionization 
potential of [\ion{S}{4}] at 34.8 eV is a factor $\sim 2$ greater than [\ion{Ar}{2}] at 
15.8 eV, [\ion{Ar}{2}] is likely weak because most of the argon is more 
highly ionized. 
This effect is seen in BCD 
galaxies, which are known to exhibit high excitation states 
\citep{2009ApJ...704.1159H, 2010ApJ...712..164H}.  As a consistency check, we note that our upper limit for 
[\ion{Ar}{3}] (ionization potential: 27.6 eV) implies a ratio of 
[\ion{Ar}{3}]/[\ion{S}{4}] that is consistent with values seen in 
local starbursts and BCDs \citep{2006ApJ...653.1129B,2006ApJ...639..157W}.

In principle, an alternative explanation for the steep continuum and 
[\ion{S}{4}] emission in J1206 is the presence of an AGN.  Indeed, Seyfert 2 
galaxies are known to show PAH and [\ion{S}{4}] features, and exhibit strong, 
rising continua due to heating of small dust grains. 
To address this concern, we consider the diagnostics of 
\citet{2000A&A...359..887L} to disentangle starburst vs. AGN energetics.  
Specifically, we use the flux ratio of 6.2\,$\mu$m PAH to $5.1-6.8$\,$\mu$m 
continuum, which is lower (higher) for a larger (smaller) AGN contribution to 
the infrared luminosity.  For J1206, this ratio is $0.66 \pm 0.08$, much 
larger than the $\lesssim 0.3$ values of AGN-dominated galaxies like NGC\,1068 
and similar to results for PDR-dominated systems like M82 and NGC\,520, which 
have $\lesssim 5 \%$ of the emission contributed by an AGN (see Figures 5 and 
6 in \citeauthor{2000A&A...359..887L}).  
Reinforcing this conclusion,  
Seyferts with similar [\ion{S}{4}] and [\ion{Ar}{2}] strengths 
have $6.2\,\mu$m PAH EWs that are $>2\sigma$
lower than for J1206 \citep{2010ApJS..187..172G}.\footnote{While such comparisons 
involve difficulties in continuum definitions (see above), we note that 
\citeauthor{2000A&A...359..887L} and \citeauthor{2010ApJS..187..172G}
also use comprehensive fits of all relevant mid-IR features to derive their 
continua, and therefore should have results similar to ours.}  Finally,  J1206 has a ratio  
[\ion{Ne}{6}]/[\ion{S}{4}]$<0.38$ ($3\sigma$), which is lower than that in any 
Seyfert for which [\ion{S}{4}] is detected \citep{2002A&A...393..821S}.  We conclude that
nuclear activity plays little role in the mid-IR spectrum of J1206.

\subsection{SDSS\,J090122.37+181432.3}
\label{sec:0901}

\begin{figure}[h]
\centering
\includegraphics[clip=true, trim=3.1cm 13.05cm 2.5cm 2.75cm,width=8cm]{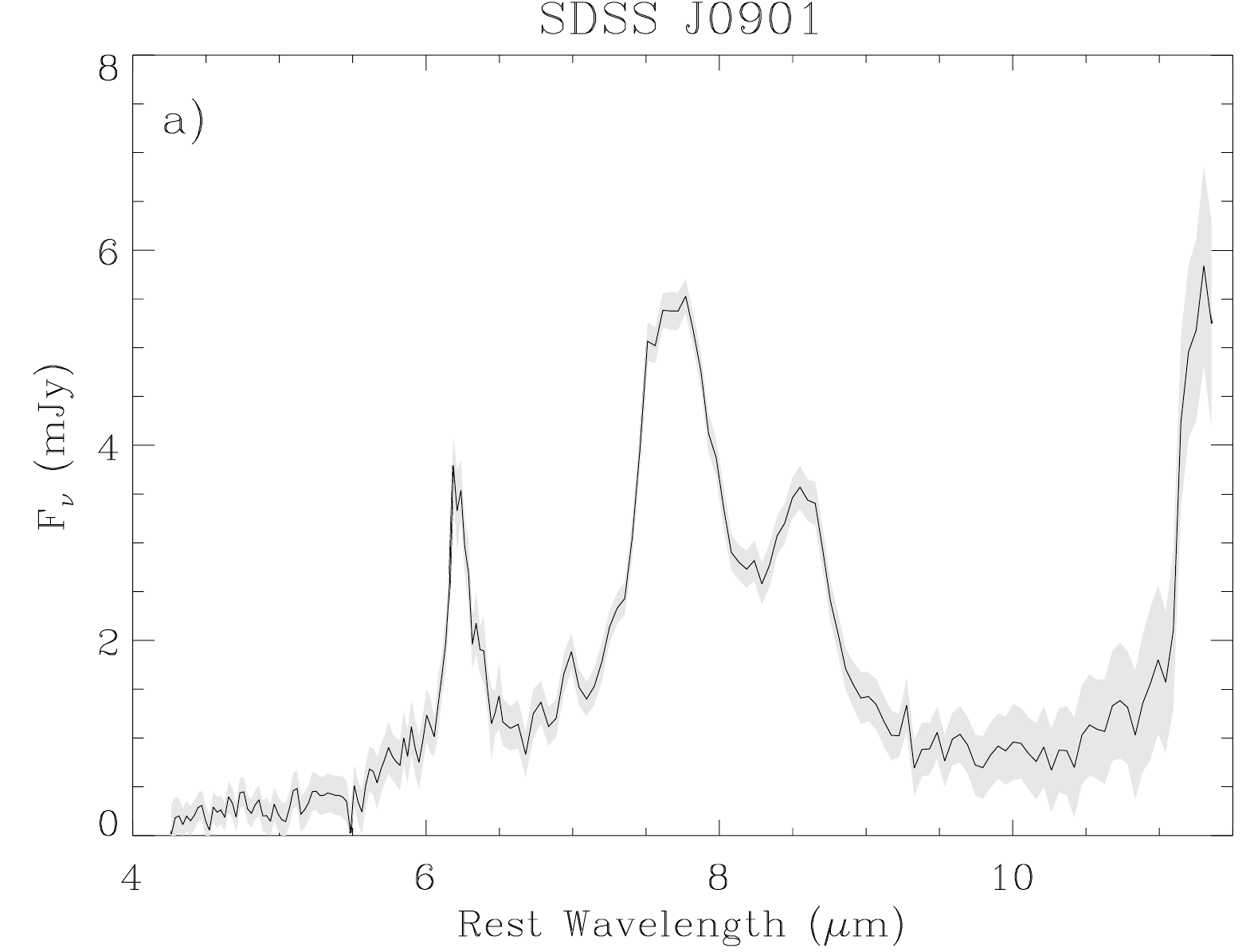}
\includegraphics[clip=true, trim=3.1cm 13.05cm 2.5cm 2.75cm,width=8cm]{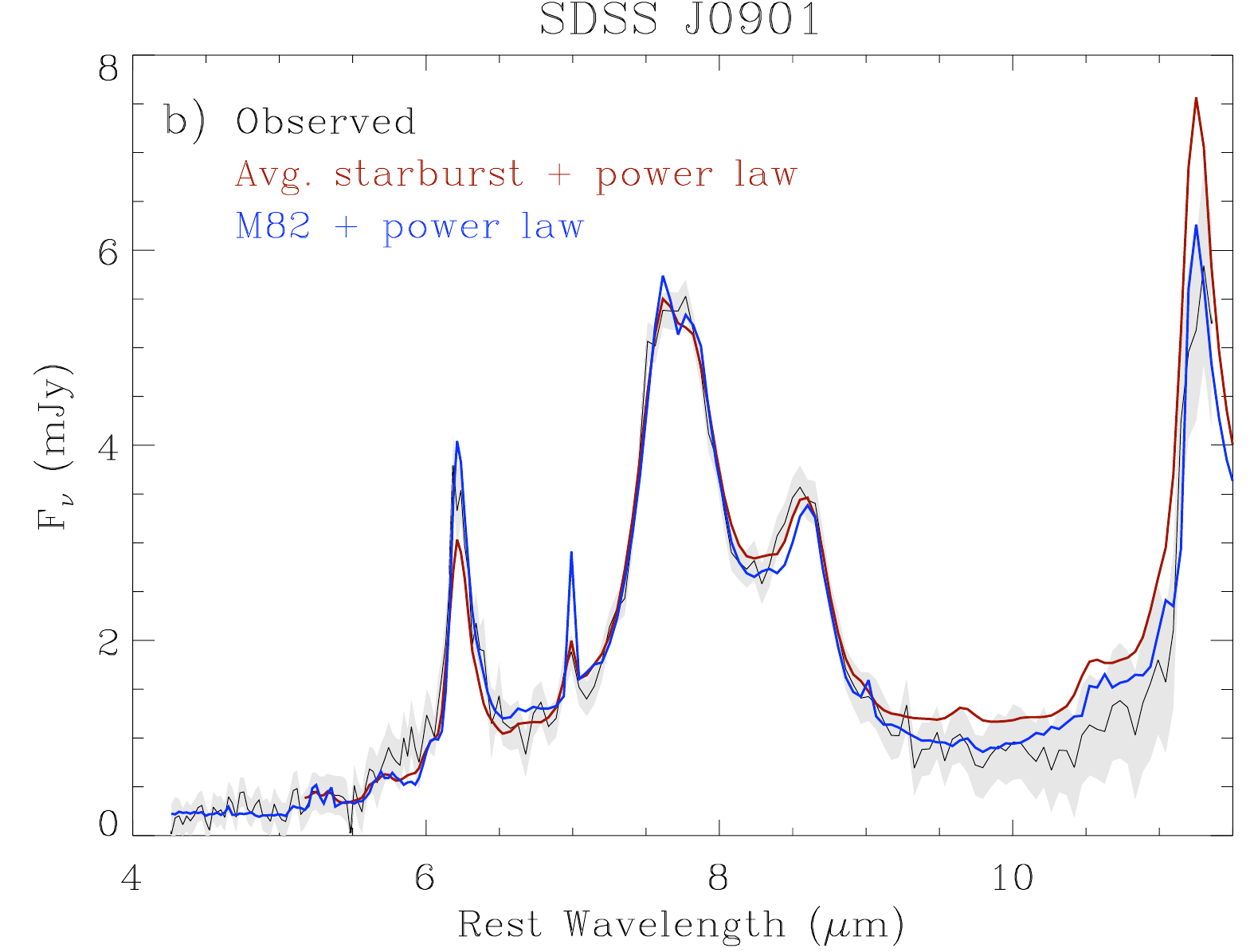}
\includegraphics[clip=true, trim=3.1cm 13.05cm 2.5cm 2.75cm,width=8cm]{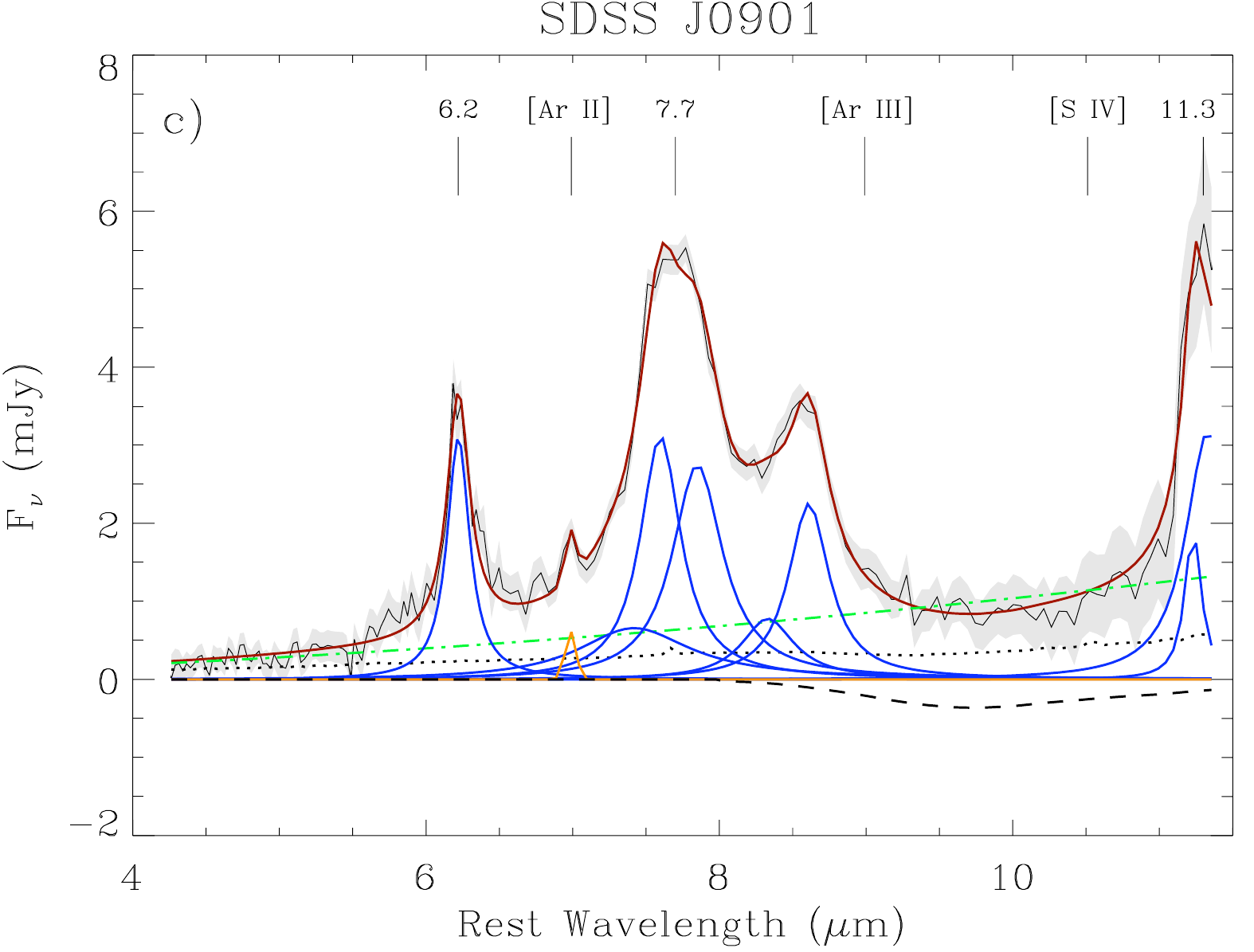}
\caption{(a) Observed IRS spectrum of J0901 plotted in black after 
redshifting to rest wavelength for $z = 2.26$. (b) Template + power law fits to the spectrum, with notation
as in Figure 2b. (c) Comprehensive fit, with notation as in Figure 2c.  The additional
dotted line indicates the AGN contribution to the spectrum implied by 
rest-frame optical measurements (see Section 3.2).}
 \label{fig:0901}
\end{figure}

Using the reduction procedure outlined in Section \ref{sec:obs}, we have 
derived the final rest-frame spectrum for J0901 that is shown in Figure 
\ref{fig:0901}a.  Following the same procedure as above, we fit the same 
starburst templates to the spectrum as in Section 3.1, with the results shown
in Table \ref{tab:templates}.  As for J1206, we find that the spectrum is best 
fit by a scaled version of M82 ($\chi_{\rm red}^2=0.95$), less well by NGC\,253 or 
the average starburst template of Brandl et al. (2006) ($\chi_{\rm red}^2=1.42, 1.26$, 
respectively), and poorly by the other Sturm et al. templates ($\chi^2_{\rm 
red} > 20$).  For the NGC\,253 and average starburst fits, the higher
$\chi^2$ values originate from the enhanced 9.7 $\mu$m silicate absorption
in J0901, which is not seen in the templates (see Figure \ref{fig:0901}b).
In contrast to J1206, J0901 exhibits only weak evidence for additional
power-law emission, with fits favoring a component that is negligible or 
has highly uncertain parameters.
Again (as in Section 3.1) exploiting the similarity to the spectrum of M82, 
and adopting a lens magnification of ${\cal M} \approx 8$ (A. West et al. 2010,
in preparation), we estimate its intrinsic far-IR luminosity to be 
$3.8 \times 10^{12}\,L_\odot$, placing it in the ULIRG regime.

As for J1206, we simultaneously fit the spectrum with a combination of 
continuum and relevant PAH features and atomic emission 
lines.  We find the degree of inferred silicate absorption is degenerate with 
the contributions of weak PAH emission features to the spectrum.  In general, 
this degeneracy has little effect on the inferred PAH emission, except in the 
relative contributions of the 11.23 and 11.33\,$\mu$m lines to the 11.3\,$\mu$m PAH feature.  In fact, the 
$6.2\,{\rm \mu m}$ and blended 7.7\,$\mu$m and 11.3\,$\mu$m PAH strengths are essentially unaffected.  
Ultimately, we have opted to use solutions with higher silicate extinction, in 
agreement with the template fits above.  We present the results of this fit 
in Table \ref{tab:results} and plot the results in Figure \ref{fig:0901}c.

The PAH emission in J0901 is strong relative 
to local starbursts, with EWs of each feature that are a factor $\sim 1.6-1.9$ times larger 
than the \citet{2006ApJ...653.1129B} average values.  The 
strengths of the 6.2, 7.7, and 11.3\,$\mu$m PAH features serve as common 
diagnostics of the 
driving mechanism(s) for infrared emission in infrared-luminous 
galaxies 
\citep[e.g.,][]{1991MNRAS.248..606R, 1998ApJ...498..579G, 2000A&A...359..887L, 2007ApJS..171...72I, 
2009ApJS..182..628V, 2010ApJ...710..289B, 2010ApJS..187..172G}.  In 
particular, suppressed PAH emission (especially at shorter wavelengths) is 
indicative of a significant AGN contribution to the bolometric infrared
luminosity.  Given the strength of the observed PAH emission, coupled 
with the shallow continuum, we conclude that accretion plays a small role 
in the mid-IR properties of J0901.  However, \citet{2009ApJ...701...52H} examined
the optical emission line ratios and found values of [\ion{O}{3}]/H$\beta$ and [\ion{N}{2}]/H$\alpha$ 
indicative of an AGN, consistent with the significant [\ion{N}{5}] and weak [\ion{Si}{4}] and 
[\ion{C}{4}] emission seen in the object's rest-UV spectrum 
\citep{2009ApJ...707..686D}.

To assess this apparent contradiction, we consider
the expected flux contribution to our mid-IR spectrum by an AGN whose
optical properties resemble those of J0901.  Specifically, we treat [\ion{O}{3}] 
5007\,\AA\, emission as a proxy for AGN strength and scale the 
$ISO$ template of NGC 1068 by the factor required to reduce the integrated [\ion{O}{3}] 
flux of NGC 1068 \citep{2006ApJS..164...81M} to the \citeauthor{2009ApJ...701...52H} value.  The dotted
line in Figure \ref{fig:0901}c shows the corresponding contribution of the scaled NGC 1068 spectrum: 
roughly 57\% and 35\% of the continuum flux at 5 and 10\,$\mu$m, respectively.
  This comparison demonstrates that reliance on rest-frame UV/optical 
measurements alone may provide a bolometrically unrepresentative picture of 
the physical properties of high-redshift systems.

Finally, we examine our IRS spectrum of J0901 for ionic and molecular 
emission.  Here J0901 is quite unlike J1206, showing [\ion{Ar}{2}] but no 
[\ion{S}{4}] emission.  The EW of [\ion{Ar}{2}] is similar to that in the 
average starburst spectrum of \citet{2006ApJ...653.1129B} (see Figure 
\ref{fig:0901}b).  As discussed above, [\ion{Ar}{2}] is weaker for systems  
with high-excitation interstellar media.  Therefore, J0901 must be bathed in 
a softer radiation field than J1206, reinforcing the conclusion that AGN 
emission plays little role in the IR properties of the system.  With our upper
limit on [\ion{Ar}{3}] we find [\ion{Ar}{3}]/[\ion{Ar}{2}]$<0.83$, which when 
combined with the argon excitation versus abundance relation from 
\citet{2003A&A...403..829V} implies 
an abundance ratio of $[{\rm Ar/H}] \gtrsim 1.3$.  Assuming that argon 
abundance is a reliable tracer of the global metallicity, this implies 
$Z \gtrsim 1.3\,Z_\odot$ for J0901.  While an intriguing result, we recommend  
caution in interpreting it given the large uncertainty (factor $\sim 2$) 
in the \citeauthor{2003A&A...403..829V} relation and its reliance on a single 
ionic line whose measurement uncertainties are $\sim25\%$.  

\section{Conclusions}
\label{sec:conclusions}

We have obtained $Spitzer$/IRS spectra of two $z\sim2$ UV-bright 
star-forming galaxies, that are 
magnified by strong gravitational lensing.  At rest wavelengths of 
$\sim5-12$\,$\mu$m, the spectra reveal strong PAH emission at 6.2, 7.7, and 
11.3\,$\mu$m, indicating that these objects are undergoing intense star formation.  
The strength of the PAH emission implies these objects have 
properties in line with those of local starbursting galaxies.  We find this 
similarity to local starburst galaxies is confirmed by our empirical template
fits, in which both galaxies are well fit by simple, rescaled versions of M82.
In detail, however, analysis of PAH strengths and emission line and continuum 
diagnostics reveals disparate properties.  We summarize our conclusions as follows:
\begin{enumerate}

\item In J1206, we find PAH EWs lower than those in the local starburst spectrum of  
\citet{2006ApJ...653.1129B}, due to an enhanced power-law continuum.  In contrast, J0901 exhibits PAH EWs that are factors 1.6--1.9 
times larger than the local average.

\item  We detect significant [\ion{S}{4}] emission in J1206.   
By analogy with two local galaxies with similar 
mid-IR spectra, NGC\,1222 and UGC\,4274, we infer a sub-solar metallicities of $\sim 0.5\,Z_\odot$, in
agreement with the published optical measurement \citep{2009ApJ...701...52H}.  
The consistency of the optical and infrared metallicity estimates suggests
J1206 has not undergone a recent violent merger.
Considering the [\ion{S}{4}]/[\ion{S}{3}] and [\ion{Ne}{3}]/[\ion{Ne}{2}] ratios 
of the local objects, we argue that J1206 is characterized by a moderately 
hard radiation field,  which naturally explains the steeply rising
continuum and lack of [\ion{Ar}{2}] emission.

\item In J0901, we detect strong PAH emission but no 
[\ion{S}{4}] or significant rising continuum.  These results indicate that the mid-IR properties of J0901 
are consistent with purely starburst-driven energetics.  This inference 
contrasts with the implications of optical spectroscopy, where emission line 
ratios show the presence of an AGN;  however, scaling from the [\ion{O}{3}] flux 
of a local AGN implies the AGN contributes $<57\%$ of the mid-IR continuum.  Thus,
from its rest-frame UV through IR properties, J0901 likely hosts a narrow-line
AGN whose IR emission is overwhelmed by that of its surrounding starburst.
This analysis highlights the need 
for future IR studies of high-redshift objects if we are to determine their 
physical properties robustly.

\item With the detection of [\ion{Ar}{2}], we are able to put an upper limit 
on the metallicity of J0901.  Using the argon abundance and excitation 
relation of \citet{2003A&A...403..829V}, we find $Z \gtrsim 1.3\,Z_\odot$, 
similar to many local starbursts.

\end{enumerate}

\acknowledgements

We thank Eckhard Sturm for providing template spectra of local galaxies 
in electronic form, and Anderson West and Tom Diehl for sharing the results 
of their lens modelling.  This work is based on observations made with the 
{\it Spitzer Space Telescope}, which is operated by the Jet Propulsion 
Laboratory, California Institute of Technology under a contract with NASA. 
Support for this work was provided by NASA through two awards issued by 
JPL/Caltech.

\bibliographystyle{apj}

\clearpage

\begin{deluxetable}{lcccc}
\tablewidth{0pt}
\tablecaption{Local starburst template fits.}
\tablehead{
Template & $\log\,(C_1)$ & $\log\,(C_2/{\rm mJy})$ & $\alpha$ & Fit type
}
\startdata
\multicolumn{5}{c}{J1206}\\
\hline
\hline
M82      &$  -5.063^{+0.006}_{-0.004} $&$ -0.78^{+0.11}_{-0.16} $&$ 3.3^{+0.2}_{-0.2} $& MCMC results\\
 &$ -5.06 $&$ -0.84 $&$ 3.1 $& Best Fit: $\chi^2_{\rm red}=0.96$ \\
\hline
NGC\,253 &$ -4.938^{+0.128}_{-0.102} $&$ -0.79^{+0.12}_{-0.14} $&$ 3.2^{+0.2}_{-0.2} $& MCMC results \\
&$ -5.03 $&$ -0.86 $&$  3.0 $& Best Fit: $\chi^2_{\rm red}=1.20$ \\

\hline
Avg. Starburst &$ -2.960^{+0.051}_{-0.051} $&$ -0.55^{+0.12}_{-0.22} $&$ 3.3^{+0.2}_{-0.2} $& MCMC results \\
&$ -2.99 $&$  -0.72 $&$ 3.1 $& Best Fit: $\chi^2_{\rm red}=1.14$ \\
\hline
\hline
\multicolumn{5}{c}{J0901}\\
\hline
\hline
M82      &$  -4.557^{+0.061}_{-0.042} $&$ -1.08^{+0.22}_{-0.16} $&$ 3.1^{+2.2}_{-2.1} $& MCMC results\\
 &$ -4.52 $&$ -1.10 $&$ 2.6 $& Best Fit: $\chi^2_{\rm red}=0.95$ \\
\hline
NGC\,253 &$ -4.675^{+0.045}_{-0.058} $&$ -0.69^{+0.13}_{-0.18} $&$ 3.2^{+0.8}_{-0.6} $& MCMC results \\
&$ -4.62 $&$ -0.56 $&$  3.6 $& Best Fit: $\chi^2_{\rm red}=1.42$ \\

\hline
Avg. Starburst &$ -2.453^{+0.038}_{-0.035} $&$ -6.51^{+0.18}_{-0.17} $&$ 0.6^{+0.1}_{-0.1} $& MCMC results \\
&$ -2.46 $&$  -6.42 $&$ 0.5 $& Best Fit: $\chi^2_{\rm red}=1.26$
\enddata
\tablecomments{$F_{\nu,{\rm fit}} = C_1 \times [{\rm 
Template}]+ C_2 \times (\lambda/6.2\,\mu{\rm m})^\alpha$.  Right-hand column indicates fit type for 
a given row.  ``MCMC results'' report the median and 68\% confidence intervals, inferred using a Monte Carlo 
Markov Chain method.  ``Best Fit'' values correspond to the highest likelihood fit, derived using optimization 
techniques.}
\label{tab:templates}
\end{deluxetable}

\footnotesize
\begin{deluxetable}{lccccc}
\tablewidth{0pt}
\tablecaption{Derived feature strengths.}
\tablehead{
 & \multicolumn{2}{c}{J1206} & \multicolumn{2}{c}{J0901} \\
Wavelength  & Observed Flux & Rest EW  & Observed Flux  & Rest EW  \\
$\mu$m & $10^{-15}$ erg s$^{-1}$ cm$^{-2}$ & $\mu$m  & $10^{-15}$ erg s$^{-2}$ cm$^{-1}$ & $\mu$m

}
\startdata
6.22 PAH & 21.7 (1.2)& 1.12 & 70.2 (3.1) & 2.12 \\
6.99 [\ion{Ar}{2}] & $<1.9$ & $<0.12$ & 3.5 (0.9) & 0.11\\
7.42 PAH & $-$ & $-$ & 52.0 (10.8)& 1.59\\
7.60 PAH & 24.2 (1.6)& 1.21 & 84.7 (4.5) & 2.61\\
7.65 [\ion{Ne}{6}] & $<0.8$ & $<0.10$ & $<2.6$ & $<0.08$ \\
7.85 PAH & 30.9 (1.8) & 1.55 & 87.0 (4.1) & 2.70\\
8.33 PAH & 3.1 (1.6) & 0.20 & 22.8 (3.9) & 0.71\\
8.61 PAH & 10.2 (1.2) & 0.51& 47.9 (2.9) & 1.51\\
8.99 [\ion{Ar}{3}] & $<2.8$ & $<0.05$ & $<2.9$ & $<0.09$ \\
10.51 [\ion{S}{4}] & 2.1 (0.1) & 0.10 & $<4.8$ & $<0.15$ \\
11.23 PAH & 3.7 (1.3) & 0.19 & 9.6 (6.6) & 0.31\\
11.33 PAH & 13.2 (3.4) & 0.61 & 42.1 (15.4)& 1.37\\
11.99 PAH & 3.2 (1.4) & 0.78 & $-$ & $-$

\enddata
\label{tab:results}
\tablecomments{Equivalent width values are based on the green dot dashed 
lines (power-law continuum fits) in Figures \ref{fig:1206}c and \ref{fig:0901}c.  Upper limits are based on $3\sigma$ uncertainties in the spectra.}
\end{deluxetable}

\end{document}